\documentclass[twocolumn,twoside,slac]{revtex4}
\usepackage{graphicx}
\usepackage{fancyhdr}
\def\ee{$e^+e^-$}
\def\to{$\rightarrow$}
\def\qq{$q\overline{q}$}

\def\WW{$WW$}
\def\ZZ{$ZZ$}

\def\ZHiggs{$Z\ Higgs$}

\def\cosTh{cos($\theta$)}
\def\cosThMax{\cosTh$_{Max}$}

\def\Pandora{Pandora~V2.2}
\def\PandoraPythia{Pandora-Pythia~V3.2}
\def\Pythia{Pythia~V6.2}
\def\Whizard{Whizard~V1.22}

\def\LC{Linear Collider}

\def\FMC{Fast Monte Carlo}
\def\LCDFMC{LCD~\FMC}
\def\SimDet{TESLA SimDet}
\def\QuickSim{JLC QuickSim}

\begin{document}

\title{Java Physics Generator and Analysis Modules}

\author{Michael T. Ronan}
\affiliation{LBNL, Berkeley, CA 94720, USA}

\begin{abstract}
A Java software framework allows modules written in different languages to be used in a high level Object-Oriented (OO) environment. Java Native Interfaces (JNI) for Linear Collider (LC) physics event generators are used in defining a common generator interface package. Portable-JNI for TESLA and Asian JLC detector simulation modules have been written for performing comparisons to the American LC detector simulation. Physics and detector Java analysis modules using prototype HEP class libraries provide high level OO study tools. Complete physics generation, parallel detector simulations and event analysis for full 500 fb$^{-1}$ simulated data samples are performed in single-pass batch jobs. Java histogram objects files are saved for final presentation using the Java Analysis Studio (JAS). The software architecture, JNI designs and overall performance is presented. Comparisons of American, Asian and European detector simulations of Higgsstrahlung events generated by Pandora, Pythia and Whizard are made.
\end{abstract}

\maketitle

\thispagestyle{fancy}

\section{INTRODUCTION}

  In performing Linear Collider physics and detector studies, one encounters a number of existing Fortran77, and new C++ and Fortran95 software modules for event generation, detector simulation and analyis. Stand-alone packages exist for Pandora(C++), Pythia(F77) and Whizard(F95) physics generation, and in some cases have been integrated into separate simulation packages for TESLA SimDet(F77), Asian JLC QuickSim(C++) and American Java Fast Monte Carlo detector studies. Individual physics analyses proceed by adopting one of the generator and detector simulation packages and often are hard to compare to similar analyses done in different environments. A Java software framework~\cite{lcstudies} allows the use of different event generators, detector simulation packages and analyis modules in the same high level Object-Oriented (OO) environment. An environment which provides modern development tools for Java physics and detector analysis modules using prototype HEP class libraries~\cite{hep.lcd}. The model allows networked graphical interactive applications, or batch single-pass processing of complete physics generation, parallel detector simulations and event analysis for full 500~fb$^{-1}$ simulated data samples. Java histogram objects files can be saved for final presentation using the Java Analysis Studio (JAS)~\cite{jas}.

  In this paper, Java Native Interfaces (JNI) for Pandora, Pythia and Whizard Linear Collider physics event generators are described and used in defining a Java framework with a common generator interface package, Sec.~\ref{sec:framework}. Sections~\ref{sec:generators} and \ref{sec:detectors} list the generator and detector simulation modules, respectively. A simple comparison of the different physics generators is made for a LC bench mark process in Sec.~\ref{sec:generator-comparisons}. Detector simulations using the U.S. Fast Monte Carlo, TESLA SimDet and Asian QuickSim packages are compared in Sec.~\ref{sec:detector-models}. Details of interface implementations for the generators, and for executing SimDet and QuickSim detector simulations and accessing simulated quantities are outlined in Sec.~\ref{sec:jni}. Java software package organization and a brief discussion of the documentation is left for Sec.~\ref{sec:packages}.

\section{JAVA ANALYSIS FRAMEWORK\label{sec:framework}}

Several prototype Java interfaces, classes and libraries have been written\cite{lcstudies} to provide a high level object-oriented physics generation, detector simulation and analysis environment for \ee\ Linear Collider studies.

\subsection{Generator interface\label{sec:interface}}
A protype Java object model has been developed to generalize the interface to physics generators such as Pandora, Pythia and Whizard. The following methods are implemented through Java Native Interfaces (JNI) to the underlying C++, Fortran77 or Fortran95 package:
\begin{itemize}
  \item setup() - Set up generator configuration using default parameters.
  \item initialize() - Initialize for selected beams, energy and physics process with given parameters.
  \item generateEvent() - Generate an event.
  \item terminate() - Summarize generated data sample.
\end{itemize}
One additional method is required to provide information about the generator.
\begin{itemize}
  \item getName() - Return generator name and version number.
\end{itemize}
Generator specific methods have been added such as
\begin{itemize}
  \item list(int) - List particles in the event.
\end{itemize}

The useage of the Java generator interface is outlined in the following excerpt:
\begin{verbatim}
// Import the Pythia generator classes.
import hep.generator.pythia.*;

// Create a Pythia process.
  pythia = new Pythia();

// Set up a "User-defined" Pythia process
// with a String of parameters settings.
  pythia.give(parameters);

// Initialize the e+e- Center-of-Mass system.
  pythia.init("CMS","e+","e-",Ecm);

// Run Pythia to generate and list events.
  for (int n=1; n<=NEvents; n++) {
    pythia.generateEvent();
    pythia.list(1);
  }
\end{verbatim}
Generator processes interfaces have been defined for classes which dynamically load predefined process settings, such as PythiaProcess for initializing Pythia. A number of reference process classes have been written, such as {\bf eetoZH} for Higgsstrahlung and {\bf eetottbar} for Top-pair production processes.

\subsection{EventGenerator and analysis modules\label{sec:modules}}
The generator interface described above is used to provide ``EventGenerator'' modules for the American LCD framework\cite{hep.lcd}. A Java package {\bf hep.lcd.generator} is abstracted to introduce different generator modules as alternative ``EventSource's'':
\begin{itemize}
  \item Generator - Defines a standard HEP physics event generator module. 
  \item GeneratedEvent - Defines a generated event.
  \item HEPEvent - Provides a standard /HEPEvt/ implementation of a generated event.
\end{itemize}
The useage of a LCD EventGenerator module is illustrated by
\begin{verbatim}
// Import the LCD generator classes for Pythia.
import hep.lcd.generator.pythia.*;

// Create a Pythia generator for given process.
  generator = new PythiaGenerator(processName);

// Run to generate and dump the first event.
  for (int n=1; n<=NEvents; n++) {
    event = generator.generateEvent();
    if (n==1) event.dump();
  }
\end{verbatim}

Within the LCD Java framework, event processing and analysis modules are added as ``Driver's'' or ``Processor's'' defined in a {\bf hep.lcd.util.driver} package.
The following classes outline the use of detector simulation and analysis modules:
\begin{verbatim}
public class AnlPythiaZHJets implements Driver

  public AnlPythiaZHJets()

    add(detectorSimulation);
    add(analysis);
 or
    add(new MCFast());
    add(new SimDetModule());
    add(new QuickSimModule());
    ...
    add(new PythiaZHJetAnalysis());

class PythiaZHJetAnalysis implements Processor
  void process(LCDEvent event)
    particles = event.get("MCParticles");
    tracks  = event.get("SimDet Tracks");
    objects = event.get("Energy Flow Objects");
    ...
\end{verbatim}
Note that the Java Virtual Machine (JVM) allows multiple detector simulations to run in independent computer memory space.

\subsection{Analysis classes and tools\label{sec:tools}}
The utility of Java physics and analysis class libraries is illustrated in the following excerpts
for the Particle physics class
\begin{verbatim}
  int PDGID = particle.getType().getPDGID();
  double[] PV = particle.getMomentum();
    double pT = Math.sqrt(PV[0]*PV[0]+PV[1]*PV[1]);
    double tanTh = pT/PV[2];
\end{verbatim}
and the Histogram analysis class
\begin{verbatim}
  histogram("Mass").fill(particle.getMass());

\end{verbatim}

Java framework applications can be built using standard HEP analysis tools  and run through
convenient Java packages so that
\begin{itemize}
  \item gmake  - Compiles Java and native code then creates a shared object library.
  \item java -mx64M GenWhizard ... - Loads Java and all shared object libraries, executes Whizard and writes a xxx.javahist unbinned histogram file.
\end{itemize}
After completing several jobs, the Java Analysis Studio~\cite{jas} can be used to combine histograms and make presentation plots.
\begin{verbatim}
  jas &
    Open files: xxx.javahist yyy.javahist ...
    Overlay and format plots
    Save as PostScript files
\end{verbatim}


\section{MONTE CARLO EVENT GENERATORS\label{sec:generators}}

Three representative LC physics generators have been used in this development: \Pandora, \Pythia\ and \Whizard.
Pythia which has been extensively tested by many experiments was used to provide checks on specific processes. Pandora and Whizard are new developments which also offer complete event generation capability.

\subsection{Pandora Monte Carlo\label{sec:pandora}}
The \Pandora~\cite{Pandora} Monte Carlo provides simulations of selected two- and three-body processes and some illustrative beyond the Standard Model processes. Initial state radiation, beam polarization, and final state spin correlations are included for all processes. In stand-alone applications Pandora is used with the \PandoraPythia~\cite{Pandora-Pythia} interface package to generate final state events. A ``straight-forward'' interface from the Java language to underlying Pandora C++ methods, Sec.~\ref{sec:jni}, is provided in a {\bf hep.generator.pythia} Java package, Sec.~\ref{sec:packages}.

\subsection{Pythia Monte Carlo\label{sec:pythia}}
The \Pythia~\cite{Pythia} Monte Carlo program allows generation of high-energy physics events including hard and soft interactions, parton distributions, initial and final state parton showers, multiple interactions, fragmentation and decay. A Java-C-Fortran77 interface described below provides full Pythia functionality within the Java environment.

\subsection{Whizard Monte Carlo\label{sec:whizard}}
The \Whizard~\cite{Whizard} Monte Carlo system is designed for the efficient calculation of multi-particle scattering cross sections and simulated event samples. Tree-level matrix elements are generated automatically for arbitrary partonic processes by calling external programs (O'MEGA, MADGRAPH and CompHEP). In this development, Java-C-Fortran95 interfaces to new Whizard procedures for initialization, phase space integration, event generation and summary was written to allow easy portability and compiler independence.

\section{DETECTOR SIMULATION MODELS\label{sec:detectors}}

Models of the expected LC detector response to 500~GeV \ee\ annihilation events are being developed. Preliminary versions used in this study are described below, and compared in Sec.~\ref{sec:detector-models}. The detailed Java interface implementations are outlined in Sec.~\ref{sec:jni}.

\subsection{LCD Fast Detector Simulation\label{sec:lcdfmc}}
The American LCD V1.4 detector simulation includes charged particle momentum smearing based on detailed error estimates, gaussian energy smearing for photons and neutral hadrons, and acceptance and energy threshold requirements. It doesn't account for confusion in reconstructing calorimeter energy deposits resulting in perfect energy flow separation of charged and neutral components of jets.

\subsection{TESLA SimDet Detector Simulation\label{sec:simdet}}
The European SimDet V4.0 detector simulation includes parameterized charged and neutral energy smearing based on full ({\bf Brahms}) Monte Carlo simulations, acceptance requirements, and a realistic energy flow algorithm.

\subsection{JLC QuickSim Detector Simulation\label{sec:quicksim}}
The Asian QuickSim detector simulation includes the generation of tracker hits followed by track and vertex reconstruction. Simulated calorimeter cell energy depositions are formed based on detailed detector simulations. Detailed cluster finding and track-cluster matching are performed before writing out combined track information..

\section{PHYSICS GENERATOR COMPARISONS\label{sec:generator-comparisons}}

\begin{figure*}[t]
\centering
 \includegraphics[width=120mm,angle=-90]{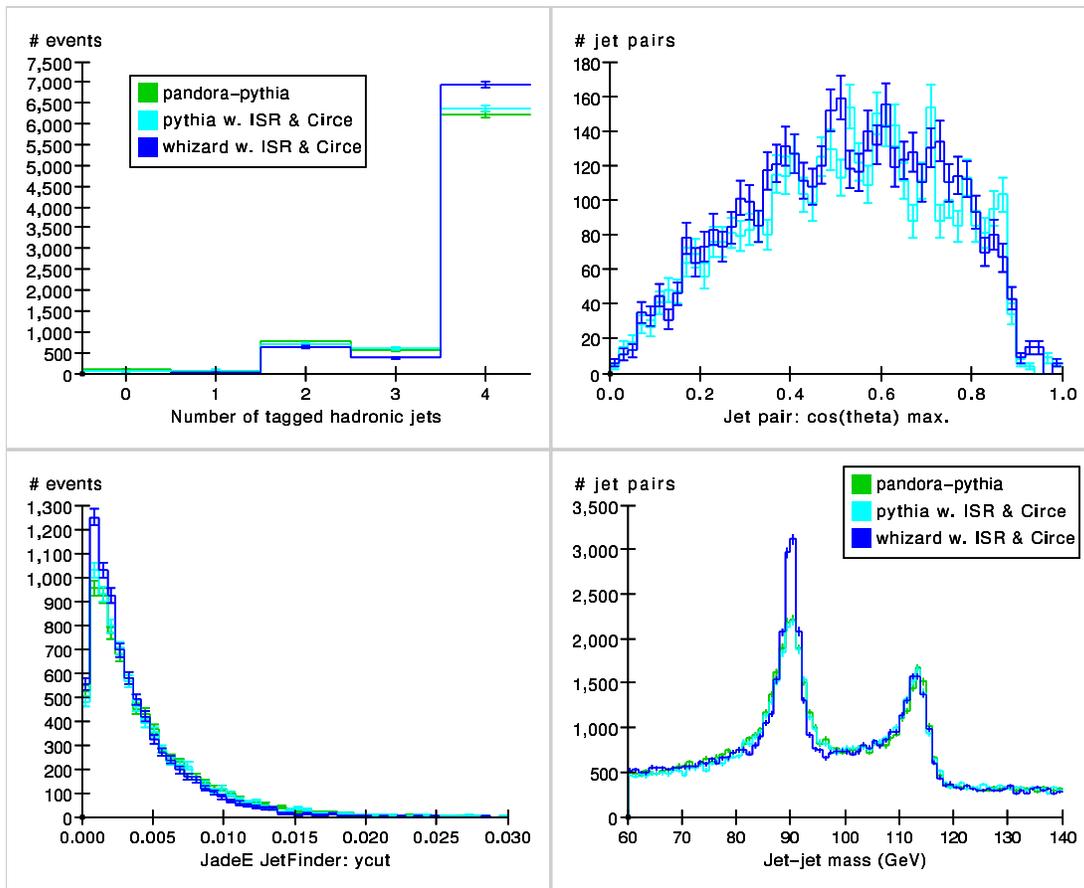}
\caption{LCD Fast Monte Carlo Particle jet distributions comparing Pandora, Pythia and Whizard generated Higgstrahlung events: a.) number of ``correctly'' reconstructed jets, b.) angular distribution (\cosThMax) of jets, c.) jet finder final ``ycut'' value, d.) jet-jet mass distributions showing Z and Higgs hadronic decay signals and the combinatorial background.}
\label{fig:jet-comparisons}
\end{figure*}

The Pandora, Pythia and Whizard Linear Collider physics event generators are compared in Fig~\ref{fig:jet-comparisons} for LCD Fast Monte Carlo detector simulations of Higgsstrahlung, \ee\to\ZHiggs, events. In these studies all generators were run at E$_{CM}$ = 500~GeV, and the Higgs mass was set to 115~GeV. Pythia hadronization was used following all parton-level event generation processes. For Pandora, the NLC500 machine parameters were selected in simulating the beamstrahlung energy spectrum. While both Pythia and Whizard event generators used Circe beamstrahlung simulation of the TESLA machine. These different machine settings are not significant in this comparison.

 Overall, the generators were found to be in good agreement. In this study, \Whizard\ generated events did not include the natural Z width as seen in the jet-jet mass distributions shown in Fig~\ref{fig:jet-comparisons}d; a Whizard deficiency that has been corrected following this work.

In related studies~\cite{higgs}, the Java framework was used to compare the simulation of background processes for \WW,\ \ZZ\ and \qq\ production from Pandora with explicit final state simulations in Whizard. These comparisons showed discrepancies due to the more precise treatment of polarization in Pandora for exclusive processes. The simulation of inclusive, interferring processes in Whizard was found to be the most complete, and the Whizard MadGraph amplitude calculation provided handelling of parton color flow information.

\section{DETECTOR COMPARISONS\label{sec:detector-models}}

\begin{figure*}[t]
\centering
 \includegraphics[width=90mm,angle=90]{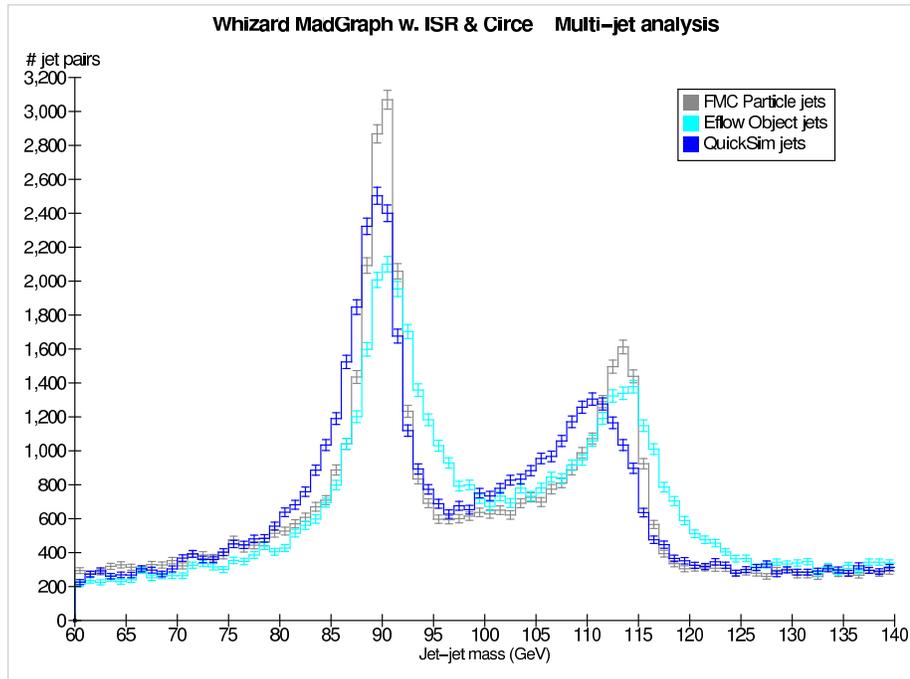}
\caption{Direct reconstruction of Z and Higgs through hadronic decays is shown for Higgsstrahlung signal events only. Jet-jet mass distributions for U.S. \LCDFMC\ (FMC), \SimDet\ and \QuickSim\ detector simulations are reconstructed for Whizard-MadGraph Monte Carlo events including ISR and Circe beamstrahlung effects.}
\label{fig:zh-comparison}
\end{figure*}

In a \ee\ \LC\ the Higgs can be reconstructed through hadronic decays\cite{higgs} even in multijet final states. The jet-jet mass distributions for the American \LCDFMC\ (FMC), \SimDet\ and \QuickSim\ detector simulations of Higgstrahlung signal-only events with the Z and Higgs decaying hadronically are shown in Fig~\ref{fig:zh-comparison}. In these current simulations, the LCD FMC jet-jet mass resolution is significantly better since it assumes ``perfect'' energy flow. More-realistic TESLA SimDet and JLC QuickSim detector simulations give comparable jet energy resolutions but different mean reconstructed jet-jet masses. Both these simulation models may be improved with the development of new techniques to resolve charged and neutral calorimeter clusters, and to correct for the loss of low energy tracks and clusters.

The overall performance of the Java framework, event generators, detector simulations and jet analyis was quite acceptable for full 500~fb$^{-1}$ production runs. Each generator was run separately with all three detector simulations being performed for the same generated event. The Java framework itself and event generation and fragmentation typically took much less than 100~msec per event, with the longest times for multi-body processes. The detailed native SimDet and QuickSim detector simulations each took about 300-500~msec per event, depending on the type of events. The Java jet-finding which included multiple passes to merge jets and the subsequent analysis took roughly 100~msec per event for each type of particle jet being found, e.g. for Monte Carlo particles, for jets found from reconstructable tracks and clusters, or for Energy Flow object jets. The typical overall event generation, multiple detector simulation, and jet analysis took about 1-1.5 seconds per event. \\
~

Batch jobs were labelled by the generator name and parameters that were set, and histogram folders were created for the different detector simulations. In the analysis of results, JAS would open separate output histogram files in different folders and use the folder names to automatically generate plot legends.

\section{JAVA NATIVE INTERFACE (JNI) IMPLEMENTATIONS\label{sec:jni}}

The physics generator, physics event, detector simulation and software framework implementations\footnote{http://www-lc.lbl.gov/software/docs/} for all developed software modules are outlined in Table~\ref{table:JNIimplementations}.

In generating events through the Pandora-JNI, a Java-C++ interface is used to allow the setting of different beam configurations, and to create physics processes to be added to Pandora's Box. Pandora processes parameters, such as the Higgs mass, are required for the Java interface construction. Once a Pandora process has been created no Java object to C++ object interface is needed. Pandora generates primary partons which are accessed through the {\bf HEPEvt-JNI} and passed to Pythia for fragmentation.

The Pythia-JNI allows for independent event generation and fragmentation processing. The interface can be used to pass configurations through pygive or to use Java classes such as ``eetoZH'' to configure Pythia for simulating the Higgsstrahlung process. The Pythia interface will be quickly replaced when new C++ versions of Pythia are made available and tested.

The Whizard interfacing to Fortran95 was simplified by creating a few simple Fortran95 modules. Whizard is configured by reading a ``whizard.in'' data file. Future developments of the Whizard Java interface are planned.

Java interfaces to the HEPEvt data structure and StdHEP output were written for multi-purpose use. After a generator fills HEPEvt, framework modules read the information and create event objects. These objects are later downloaded into HEPEvt structures that are accessed by fragmentation processes or by the different detector simulations. The design allows any combination of physics generator, fragmentation process or detector simulation. User analysis interfaces have also been written to allow one to use existing analysis packages.

An interface to Circe~\cite{Circe} is used to handle the initialization of beamstrahlung simulation parameters from the Java framework for the Pythia and Whizard generators. The Tauola~\cite{Tauola} $\tau$-decay package was used to handle polarization effects. A simple Java interface was written to allow testing of individual Tau decay modes.

The detector simulation packages have simple interfaces to set up, initialize and execute the simulations. Care was taken to pass events into the packages through separate HEPEvt interfaces. The simulated detector responses for SimDet and QuickSim are accessed from underlying storage into Java track and cluster classes to form reconstructed objects that are added to the LCD event for subsequent analysis. The SimDet and QuickSim interfaces follow a similar design model which could be generalized for other applications.

\begin{table*}[t]
\begin{center}
\caption{Java Native Interface (JNI) Implementations}
\begin{tabular}{|l|l|l|l|}
\hline \textbf{Package} & \textbf{Java method signature} & \textbf{Native invocation} \\
\hline \underline{Physics Generators} &  &  \\
 \textbf{Pandora} &  &  \textbf{C++ $\rightarrow$ C++} \\
 \hspace{2mm} Create a new beam. & native newBeam(Eb,pol,in,out); & b1 = new ebeam(Eb,pol,in,out); \\
 \hspace{2mm} Create a new process. & native newProcess(name,param); & P = new processName(param); \\
 \hspace{2mm} Add process to Pandora's box. & native add(PandoraProcess process); & Box$\rightarrow$add(*P); \\
 \hspace{2mm} Generate an event.          & native nextEvent();      & LEvent LE = Box$\rightarrow$getEvent(); \\
 &  &  \\
\hline \textbf{Pythia} &  & \textbf{C $\rightarrow$ F77} \\
 \hspace{2mm} Give a parameter list.      & give(String s) $\{$ give(s,s.length); $\}$ &  \\
 \hspace{2mm}                             & native give(String s,int len);  & pygive$\_$(string,length); \\
 \hspace{2mm} Initialize CMS frame.       & native init(String frame, ...); & pyinit$\_$(frame,b1,b2,$\&$E,i,j,k); \\
 \hspace{2mm} Generate an event.          & native exec();      & pyexec$\_$(); \\
 &  &  \\
\hline \textbf{Whizard} &  & \textbf{C $\rightarrow$ F95} \\
 \hspace{2mm} Read input data file.       & native readInput(); & whizard$\_$read$\_$input$\_$(); \\
 \hspace{2mm} Do phase space integration. & native integrate(); & whizard$\_$integrate$\_$(); \\
 \hspace{2mm} Generate an event.          & native event();     & whizard$\_$event$\_$(); \\
 &  &  \\
\hline \textbf{Circe}   &  & \textbf{C $\rightarrow$ F77} \\
 \hspace{2mm} Initialize beamstrahlung simulation. & native init(double X1M, ...); & circes$\_$($\&$x1m, ...); \\
 &  &  \\
\hline \textbf{Tauola}  &  & \textbf{C $\rightarrow$ F77} \\
 \hspace{2mm} Set decay mode of Tau+ or Tau-. & native setDecayMode(int i, int mode); & if (i==1) jaki$\_$.jak1 = mode; \\
                                              &                                       & else if (i==2) jaki$\_$.jak2 = mode; \\
 &  &  \\
\hline
\hline \underline{Physics Events} &  &  \\
\hline \textbf{HEPEvt}  &  & \textbf{C $\rightarrow$ F77} \\
 \hspace{2mm} Get momentum vector component.  & native getPHEP(int i, int j); & return hepevt$\_$.phep[i][j]; \\
 &  &  \\
\hline \textbf{StdHEP}  &  & \textbf{C $\rightarrow$ F77} \\
 \hspace{2mm} Initialize output file. & native initialize(String outfile, ...); & stdhinit$\_$(outfile,$\&$ngen); \\
 \hspace{2mm} Write an event.         & native writeEvt(); & stdhwrit$\_$($\&$nevt); \\
 &  &  \\
\hline
\hline \underline{Detector Simulations} &  &  \\
 \textbf{SimDet} &  & \textbf{C $\rightarrow$ F77} \\
 \hspace{2mm} Initialize detector model. & native init(); & siinit$\_$(); \\
 \hspace{2mm} Set detector parameters.   & native detr(); & sidetr$\_$(); \\
 \hspace{2mm} Simulate detector response & native exec(); & siexec$\_$(); \\
 &  &  \\
\hline \textbf{QuickSim} &  & \textbf{C++ $\rightarrow$ C++/F77} \\
 \hspace{2mm} Create a JSF package.  .   & native newJSF(); & jsf  = new JSFSteer(...); \\
 \hspace{2mm} Set up QuickSim process.   & native setup();  & sim = new JSFQuickSim(...); \\
 \hspace{4mm}   Set type of generator. & native setGenerator(type);  & sim$\rightarrow$setGenerator(type); \\
 \hspace{2mm} Simulate detector response & native exec();   & result = jsf$\rightarrow$Process(eventNum); \\
 &  &  \\
\hline
\hline \underline{Software Frameworks} &  &  \\
 \textbf{JSF} &  & \textbf{C++ $\rightarrow$ C++} \\
 \hspace{2mm} Create a new JSF framework. & native newJSF(); & jsf  = new JSFSteer(...); \\
 \hspace{2mm} Add a LCFull process. & native addLCFull(); & full = new JSFLCFULL(...); \\
 \hspace{2mm} Set up defaults: & native setup();  & sim = new JSFQuickSim(...); \\
 \hspace{4mm}   Set type of generator. & native setGenerator(type);  & sim$\rightarrow$setGenerator(type); \\
 \hspace{2mm} Add an analysis process. & native addAnalysis(); & anal = new MyAnalysis(); \\
 \hspace{2mm} ... &  &  \\
\hline
\end{tabular}
\label{table:JNIimplementations}
\end{center}
\end{table*}

\section{JAVA SOFTWARE PACKAGE AND API ORGANIZATION\label{sec:packages}}

The interface software and LCD modules have been organized into several Java packages or class libraries. The interfaces to the imported physics generator and detector simulation modules are in the HEP packages listed in Table~\ref{table:HEPpackages}. The LCD framework modules are organized in the packages listed in Table~\ref{table:LCDpackages}. The LCD modules provide reference implementations using the more generic interfaces to the HEP packages.

\begin{table*}[t]
\begin{center}
\caption{HEP Java software package}
\begin{tabular}{|l|l|l|}
\hline \textbf{Project} & \textbf{Java classes} & \textbf{Java package name} \\
\hline \underline{HEP Generator packages} &  & hep.generator \\
 Pandora & Pandora, PandorasBox, Beam,             & hep.generator.pandora \\
         & PandoraProcess, eetoee, eetoZHiggs, ... &  \\
         &                 &  \\
 Pythia  & Pythia, PythiaProcess, eetobbbar,       & hep.generator.pythia  \\
         & eetoZH, ...     &  \\
         &      &  \\
 Whizard &  & hep.generator.whizard \\
  &      &  \\
\underline{also}     &  &  \\
 Circe   & Circe  & hep.generator.circe   \\
  &      &  \\
 Tauola  & Tauola & hep.generator.tauola  \\
  &  &  \\
\hline
\hline \underline{HEP Event package} &  &  \\
 HEP     & HEPEvt, StdHEP  & hep.event      \\
  &  &  \\
\hline
\hline \underline{TESLA packages} &  &  \\
 SimDet detector simulation   & SimDet, SimDetCHAPAEntry, & hep.tesla.simdet \\
                              & SimDetCLUSEntry, SimDetEFLOWEntry, ...     &  \\
  &  &  \\
\hline
\hline \underline{JLC} &  &  \\
 JSF Software framework       & JSFSoftwareFramework, JSFGenerator,        & hep.jlc.jsf \\
                              & JSFGeneratorParticle, JSFPythiaGenerator   &  \\
  &      &  \\
\hline
 QuickSim detector simulation & JSFPackage, QuickSim,                      & hep.jlc.quicksim \\
                              & QuickSimParticleEntry, QuickSimTrackEntry, &  \\
                              & QuickSimCalHitEntry, ... &  \\
  &  &  \\
\hline
\end{tabular}
\label{table:HEPpackages}
\end{center}
\end{table*}

The LCD event generator and detector simulation modules are designed for simple use, see Sec.~\ref{sec:modules}. The Generator, Fragmentation and GeneratedEvent interfaces allow the design of abstract models for sourcing events to be based to the different detector simulation modules. Simulated reconstructed objects implement a basic Particle class allowing the development of abstract analysis methods to reconstruct event quantities from different detector models.

\begin{table*}[t]
\begin{center}
\caption{LCD Java software package organization}
\begin{tabular}{|l|l|l|}
\hline \textbf{Package} & \textbf{Java classes} & \textbf{Java package name} \\
\hline \underline{Generator packages} &  &  \\
 Physics generators & Generator, Fragmentation & hep.lcd.generator \\
                    & GeneratedEvent, HEPEvent &  \\
  &  &  \\
 Pandora generator  & PandoraGenerator, PandoraEvent, & hep.lcd.generator.pandora \\
                    & LCDPandoraEvent &  \\
  &  &  \\
 Pythia generator   & PythiaGenerator, PythiaFragmentation, & hep.lcd.generator.pythia \\
                    & PythiaEvent, LCDPythiaEvent     &  \\
  &  &  \\
 Whizard generator  & WhizardGenerator, WhizardEvent, & hep.lcd.generator.whizard \\
                    & LCDWhizardEvent &  \\
  &  &  \\
\hline
\hline \underline{Monte Carlo packages} &  &  \\
  &  & hep.lcd.mc \\
  &  &  \\
 U.S. Fast MC Detector Simulation & MCFast, ReconTrack, & hep.lcd.mc.fast \\
                                  & ReconCluster, ...   &  \\
  &  &  \\
 TESLA SimDet Detector Simulation & SimDetModule, SimDetDriver, & hep.lcd.mc.simdet \\
                                  & SimDetAdapter, SimDetParticle,  &  \\
                                  & SimDetTrackList, SimDetClusterList, ... &  \\
  &  &  \\
 JLC QuickSim Detector Simulation & QuickSimModule, QuickSimDriver, & hep.lcd.mc.quicksim \\
                                  & QuickSimAdapter, QuickSimParticle,  &  \\
                                  & QuickSimTrackList, QuickSimClusterList, ... &  \\
  &  &  \\
\hline
\hline \underline{Software framework packages} &  &  \\
  &  & hep.lcd.framework \\
  &  &  \\
 JLC Software Framework           & JLCSoftwareFrameworkDriver, JSFEventGenerator, & hep.lcd.framework.jsf \\
                                  & JSFGeneratorParticle, ... &  \\
  &  &  \\
\hline
\end{tabular}
\label{table:LCDpackages}
\end{center}
\end{table*}

Special utility packages, {\bf pandorapythia} and {\bf whizardpythia}, based on the Pandora-Pythia~\cite{Pandora-Pythia} software model for handelling the physics interfacing from Pandora and Whizard generated parton color flow information to Pythia hadronization, are located in the {\bf hep.lcd.util} package.

Software Application Programming Interface (API) documentation on all Java packages is available on the web\footnote{http://www-lc.lbl.gov/software/java/api/}. Class inheritance and method signiture documentation is automatically generated. Some documentation exists for most classes while comments on many self-explanatory methods are missing. \\
~

In building stand-alone event generation, detector simulation and physics analysis packages, the Java classes are compiled according to the following rule

\begin{verbatim}
# For a collection of Java source files
 JAVAFILES := $(wildcard hep/.../*.java)

# the corresponding class files are obtained
 CLASSFILES:= $(patsubst %.java, \
                %.class,$(JAVAFILES))

# and compiled with
%.class: %.java
   @echo "Compiling" $(*F)
   @javac $(JAVACFLAGS) $<
\end{verbatim}
Header files for the JNI implementations are automatically generated from the Java source code using a utility
\begin{verbatim}
$(HeaderFile): $(JNISource)
   @if [ -f $(TargetFile) ]; \
      then rm $(TargetFile); fi;
   @javah -jni -d $(TargetDir) $(Source)
\end{verbatim}
Shared object libraries and native code are compiled following
\begin{verbatim}
   gcc -shared -I$(JavaJDK)/include \
     -I$(JavaJDK)/include/linux \
     ...
     -L/usr/lib/gcc-lib/i386-redhat-linux/2.96 -lg2c \
     -L/usr/lib -lc \
     -o $(SharedLib)
\end{verbatim}
where JavaJDK locates the Java development kit (JDK), the include files are used by the JNI implementations, and the library list usually includes Linux specific libraries and general libraries that normally would be loaded into an executable image. The shared object library is located by the LD\_LIBRARY\_PATH environment variable and loaded into the JavaVM following
\begin{verbatim}
public class Pythia
{
  ...

  /** Static initializer loads Pythia library. */
  static {
    String libname = "pythia";
    System.out.println(" Load lib"+libname+".so");
    System.loadLibrary(libname);
  }
}
\end{verbatim}

\newpage

\begin{acknowledgments}
I thank Tony Johnson for many useful discussions on Java object modeling and for presenting this paper at the conference. Tony provided an initial interface to the HEPEvt data structure. Michael Peskin and Tim Barklow provided guidance on interfacing to Pandora and on passing the parton color flow information to Pythia for fragmenting high-energy \ee\ jets. The Pythia generator interface was extended to include pygive \& pyinit in consultation w/ Stephen Mrenna. Tim Barklow, Wolfgang Kilian and Thorsten Ohl helped in defining the Whizard interface and in implementing simple Fortran95 procedures. I thank Heinz-J\"{u}rgen Schreiber and Akiya Miyamoto for reviewing the detailed interfacing to the TESLA SimDet and JLC QuickSim detector simulation packages. The Java physics generator and detector simulation modules described in this paper are used within the Java based software framework under development by the American LCD software group\cite{lcstudies,hep.lcd,jas}.

Work supported by the U.S. Department of Energy contract DE-AC03-76SF00098.
\end{acknowledgments}


\begin{thebibliography}{9}   

\bibitem{lcstudies} All modules and tools used in this study are available from
      {\bf http://www.lbl.gov/ $\sim$ronan/docs/lcstudies}.

\bibitem{hep.lcd} Norman Graf et al, Proceedings of the International Workshop on Linear Colliders,
Jeju Island, Korea, August 26-30, 2002 and references therein.

\bibitem{jas} Tony Johnson, SLAC, {\it Java Analysis Studio}, {\bf http://jas.freehep.org}.

\bibitem{Pandora} PANDORA~V2.2, {\bf http://www-sldnt.slac. stanford.edu/nld/new/Docs/Generators/ PANDORA.htm}.

\bibitem{Pandora-Pythia} PANDORA-PYTHIA~V3.2, {\bf http://www-sldnt.slac.stanford.edu/nld/new/Docs/ Generators/PANDORA\_PYTHIA.htm}

\bibitem{Pythia} PYTHIA~V6.2, {\bf http://www.thep.lu.se/ $\sim$torbjorn/Pythia.html}.

\bibitem{Whizard} WHIZARD~V1.22, {\bf http://www-ttp.physik. uni-karlsruhe.de/Progdata/whizard/}.

\bibitem{Circe} CIRCE~V7.0, {\bf http://heplix.ikp.physik.tu-darmstadt.de/lc/beam.html}.

\bibitem{Tauola} TAUOLA~V2.6, {\bf http://hpjmiady.ifj.edu.pl/ programs/node6.html}.

\bibitem{SimDet} SIMDET~V4.0, {\bf http://www-zeuthen. desy.de/linear\_collider/}.

\bibitem{JSF} JSF~V2002-1.0, {\bf http://www-jlc.kek.jp/ subg/offl/jsf/docs/usersguide/html/}.

\bibitem{QuickSim} QUICKSIM is part of LCLIB, {\bf http://www-jlc.kek.jp/subg/offl/lclib/index.html}.

\bibitem{higgs} Michael T. Ronan, {\it Multi-Jet Higgsstrahlung Event Analysis''}, LCWS 2002, ibid.

\end{thebibliography}

\end{document}